\documentclass[preprintnumbers,superscriptaddress,showpacs,pra]{revtex4}%
\usepackage[colorlinks=true,linkcolor=blue]{hyperref}
\usepackage{amssymb}
\usepackage{amsfonts}
\usepackage{amsmath}
\usepackage{graphicx}%
\providecommand{\U}[1]{\protect\rule{.1in}{.1in}}

\begin{document}
\preprint{REV\TeX4-1 }
\title[Geometric Momentum on Sphere]{Geometric momentum in the Monge parametrization of two dimensional sphere}
\author{D. M. Xun, and Q. H. Liu}
\email{quanhuiliu@gmail.com}
\affiliation{School for Theoretical Physics, and Department of Applied Physics, Hunan
University, Changsha, 410082, China}

\begin{abstract}
A two dimensional surface can be considered as three dimensional shell whose
thickness is negligible in comparison with the dimension of the whole system.
The quantum mechanics on surface can be first formulated in the bulk and the
limit of vanishing thickness is then taken. The gradient operator and the
Laplace operator originally defined in bulk converges to the geometric ones on
the surface, and the so-called geometric momentum and geometric potential are
obtained. On the surface of two dimensional sphere the geometric momentum in
the Monge parametrization is explicitly explored. Dirac's theory on
second-class constrained motion is resorted to for accounting for the
commutator $\left[  x_{i},p_{j}\right]  =i\hbar\left(  \delta_{ij}-x_{i}%
x_{j}/r^{2}\right)  $ \ rather than $\left[  x_{i},p_{j}\right]  =i\hbar
\delta_{ij}$ \ that does not hold true any more. This geometric momentum is
geometric invariant under parameters transformation, and self-adjoint.

\end{abstract}
\date{\today}

\maketitle

\section{Introduction}

It has been a long standing problem how to properly define quantum mechanics
on a surface in three dimensional Euclidean space. On one hand, Dirac stressed
that in his \emph{Principles} on the canonical quantization assumption that
"is found in practice successful only when applied with the dynamic
coordinates and momenta referring to a Cartesian system of axes and not to
more general curvilinear coordinates." \cite{dirac1} On the other hand, there
is in textbooks a routine recipe proposed by DeWitt by hypothesizing the
quantum kinetic energy operator to be proportional to Laplace-Beltrami
operator $\Delta_{LB}$ on the surface, \cite{dewitt}%
\begin{equation}
T=-\frac{\hbar^{2}}{2m}\Delta_{LB}.
\end{equation}
Is there a way to start from Dirac to reach DeWitt? Certainly, Dirac put
forward a theory for systems of second-class constraints \cite{dirac2}\ which
really encompasses the DeWitt's hypothesis as a special case, \cite{1990} but
contains much more than what was expected.

When we use the tensor covariant and contravariant components and the Einstein
summation convention, the so-called standard parametrization $\mathbf{r}%
(q^{1},q^{2})$ of the 2D surface is given by,
\begin{equation}
\mathbf{r}(q^{1},q^{2})\mathbf{=(}\text{ }x(q^{1},q^{2}),y(q^{1}%
,q^{2}),z(q^{1},q^{2})\text{ }\mathbf{).} \label{standform}%
\end{equation}
In differential geometry, ($q^{1},q^{2}$) is generally denoted by $q^{\mu}$
and $q^{\nu}$ with lowercase greek letters $\mu,\nu$ taking values $1,2$, and
$\mathbf{r}^{\mu}=g^{\mu\nu}\mathbf{r}_{\nu}=g^{\mu\nu}{\partial_{\nu}%
}\mathbf{r}$ $=g^{\mu\nu}\partial\mathbf{r/}q^{\nu}$ with $g_{\mu\nu
}={\partial_{\mu}}\mathbf{r}\cdot{\partial_{\nu}}\mathbf{r}$ being the metric
tensor. At this point $\mathbf{r}$, $\mathbf{n=(}n_{x},n_{y},n_{z}\mathbf{)}$
is the normal and $M\mathbf{n}$ symbolizes the mean curvature vector field, a
geometric invariant. \cite{liu07} In physics, this two dimensional (2D)
surface can more realistically be considered as a 3D shell whose thickness is
negligible in comparison with the dimension of the whole system. Then, there
are two ways to performing the calculus on the surface: Explicitly, when the
2D curved surface is conceived as a limiting case of a curved shell of equal
thickness $d$, where the limit $d\rightarrow0$ is then taken, great
discrepancies present as firstly taking limit $d\rightarrow0$ then defining
the derivatives on the surface, and as firstly defining derivatives in bulk
then letting $d\rightarrow0$. The second order is named as the \emph{confining
}$\emph{procedure}$ for studying motion on 2D surface embedded in 3D.
\cite{jk,dacosta,FC,BP} This kind of exploration was initialized in 1971,
\cite{jk} fundamentally finished in 1981, \cite{dacosta} and with correct
inclusion of electromagnetic field in 2008 \cite{FC} etc. Remarkably, as the
\emph{confining procedure} is applied to the momentum operator $\mathbf{p}%
=-i\hbar\nabla$, we find that the resultant momentum on the surface is with
$M$ denoting the mean curvature, \cite{liuPRA,liu11}
\begin{equation}
\mathbf{p}=-i\hbar(\mathbf{r}^{\mu}\partial_{\mu}+M\mathbf{n}), \label{GM}%
\end{equation}
which was originally discovered in 2007 \cite{liu07} by an entirely
independent development on the quantization of the momentum on 2D surface
embedded in 3D flat space. This momentum corresponds to the so-called standard
parametrization $\mathbf{r}(q^{1},q^{2})$ of the 2D surface (\ref{standform})
in mathematics therefore should be preferable over other forms of momentum
such as the generalized momenta ($p_{q^{1}},p_{q^{2}}$) canonically conjugated
to parameters ($q^{1},q^{2}$). Paralleling to the confining procedure-induced
\emph{geometric potential } $V_{gp}\equiv-\hbar^{2}/2m(M^{2}-K)$ with $K$
being the gaussian curvature, \cite{Szameit,Schultheiss} we call (\ref{GM})
\emph{geometric momentum}. \cite{liu11} This scheme of building up quantum
mechanics on the surface echoes the historic comments of Dirac on the
canonical quantization in his \emph{Principles}. \cite{dirac1}

In 2010, with help of the femtosecond laser writing technology, the optical
analogue of the quantum \emph{geometric potential} is experimentally realized
and its experimental effects on optical wave packets constrained on curved
surfaces are demonstrated. \cite{Szameit} In 2012, the \emph{geometric
potential} effects on the electronic properties of materials such as
Tomononaga-Luttinger liquids are directly confirmed with an observation of the
\textit{in situ} high-resolution ultraviolet photoemission spectra of a
one-dimensional metallic $C_{60}$ polymer with an uneven periodic
peanut-shaped structure. \cite{epl} These two experimental verifications may
have influences on further developments of physics and mathematics for the 2D
curved surfaces, for the \emph{geometric momentum }and the\emph{ geometric
potential }are, upon two constant factors, the gradient and Laplacian
operator, respectively, as pointed out in Refs. \cite{liuPRA,liu11}.

The principal purpose of this study is to explicitly show that, with use of
the Monge parametrization of the 2D surface, the geometric momentum is
compatible with Dirac's theory for systems of second-class constraints all around.

\section{Geometric momentum with Cartesian variables ($x,y$)}

By Monge parametrization, we mean that a 2D surface given by the form
$z=f(x,y)$ where ($x,y,z$) are Cartesian variables. For a sphere of radius $r$
in $R^{3}$, we have the so-called standard form,%
\begin{equation}
\mathbf{r}(x,y)=(x,y,\sqrt{r^{2}-x^{2}-y^{2}}).
\end{equation}
The covariant derivatives $\mathbf{r}_{\mu}$ and contravariant derivatives
$\mathbf{r}^{\mu}$ can be easily computed and the results are respectively,%
\begin{equation}
\left(
\begin{array}
[c]{c}%
\mathbf{r}_{_{x}}\\
\mathbf{r}_{_{y}}%
\end{array}
\right)  =\left(
\begin{array}
[c]{lll}%
1, & 0, & -x/\sqrt{r^{2}-x^{2}-y^{2}}\\
0, & 1, & -y/\sqrt{r^{2}-x^{2}-y^{2}}%
\end{array}
\right)  ,
\end{equation}%
\begin{equation}
\left(
\begin{array}
[c]{c}%
\mathbf{r}^{x}\\
\mathbf{r}^{y}%
\end{array}
\right)  \equiv\left(
\begin{array}
[c]{c}%
g^{x\mu}\mathbf{r}_{\mu}\\
g^{y\mu}\mathbf{r}_{\mu}%
\end{array}
\right)  =\frac{1}{r^{2}}\left(
\begin{array}
[c]{lll}%
r^{2}-x^{2}, & -xy, & -x\sqrt{r^{2}-x^{2}-y^{2}}\\
-xy, & r^{2}-y^{2}, & -y\sqrt{r^{2}-x^{2}-y^{2}}%
\end{array}
\right)  .
\end{equation}
The normal $\mathbf{n}$ and the mean curvature $M$ are given by respectively,%
\begin{equation}
\mathbf{n=}\frac{1}{r}(x,y,\sqrt{r^{2}-x^{2}-y^{2}}),\text{ }M=-\frac{1}{r}.
\end{equation}
Then, the geometric momentum operators $p_{i}$ $(i=x,y,z)$ are,%
\begin{align}
p_{x}  &  =-i\hbar\frac{1}{r^{2}}(\left(  r^{2}-x^{2}\right)  \frac{\partial
}{\partial x}-xy\frac{\partial}{\partial y}-x),\label{px}\\
p_{y}  &  =-i\hbar\frac{1}{r^{2}}(-xy\frac{\partial}{\partial x}+\left(
r^{2}-y^{2}\right)  \frac{\partial}{\partial y}-y),\label{py}\\
p_{z}  &  =i\hbar\frac{\sqrt{r^{2}-x^{2}-y^{2}}}{r^{2}}(x\frac{\partial
}{\partial x}+y\frac{\partial}{\partial y}+1). \label{pz}%
\end{align}
In flat space, we take commutator $[x_{i},p_{j}]=i\hbar\delta_{ij}$ for
granted. But we can easily verify that one the sphere the correct results turn
out to be $\left[  x_{i},p_{j}\right]  =i\hbar\left(  \delta_{ij}-x_{i}%
x_{j}/r^{2}\right)  $ with use of (\ref{px})- (\ref{pz}). In next section, we
will show that we need the Dirac's theory for systems of second-class
constraints which accounts for this fact.

\section{Dirac's theory for systems of second-class constraints}

On the sphere in the Monge paramerization, the primary Hamiltonian $H_{p}$ is,
\cite{dirac2}%
\begin{equation}
H_{p}=\frac{p_{i}^{2}}{2m}+\lambda\left(  \sqrt{r^{2}-x^{2}-y^{2}}-z\right)
+up_{\lambda}, \label{ham}%
\end{equation}
where $\lambda$ is the Lagrangian multiplier enforcing the constrained of
motion on the surface, and $u$ is also a Lagrangian multiplier guaranteeing
that this Hamiltonian is defined on the symplectic manifold, and $p_{i}$
$\left(  i=x,y,z\right)  $ and $p_{\lambda}$ are respectively the canonical
momenta conjugate to variables $x_{i}$\ and $\lambda$. The Poisson bracket is
defined by.
\begin{equation}
\left\{  f,H_{p}\right\}  \equiv\frac{\partial f}{\partial x_{i}}%
\frac{\partial H_{p}}{\partial p_{i}}+\frac{\partial f}{\partial\lambda}%
\frac{\partial H_{p}}{\partial p_{\lambda}}-(\frac{\partial f}{\partial p_{i}%
}\frac{\partial H_{p}}{\partial x_{i}}+\frac{\partial f}{\partial p_{\lambda}%
}\frac{\partial H_{p}}{\partial\lambda}). \label{poisson}%
\end{equation}
The equations of motion for ($x,y,z,\lambda$) are given by,%
\begin{equation}
p_{i}=m\dot{x},\text{ }p_{\lambda}=0. \label{plamb}%
\end{equation}
The primary constraint is then,%
\begin{equation}
\varphi_{1}=p_{\lambda}\approx0,
\end{equation}
hereafter symbol "$\approx$" implies a weak equality. After all calculations
are finished, weak equality takes back the strong one. The secondary
constraints (not confusing with second-class constraints) are then determined
by,%
\begin{equation}
\left\{  \varphi_{i},H_{p}\right\}  \approx0.
\end{equation}
And the complete secondary constraints are,%
\begin{align}
\varphi_{2}  &  =\sqrt{r^{2}-x^{2}-y^{2}}-z\approx0,\label{fir}\\
\varphi_{3}  &  =\frac{xp_{x}+yp_{y}}{m\sqrt{r^{2}-x^{2}-y^{2}}}+\frac{p_{z}%
}{m}\approx0,\label{sec}\\
\varphi_{4}  &  =\frac{\left(  r^{2}-x^{2}-y^{2}\right)  (p_{x}^{2}+p_{y}%
^{2}+p_{z}^{2})-r^{2}\sqrt{r^{2}-x^{2}-y^{2}}m\lambda}{m^{2}\left(
r^{2}-x^{2}-y^{2}\right)  ^{3/2}}\approx0,\label{thi}\\
\varphi_{5}  &  =\frac{p_{z}(p_{x}^{2}+p_{y}^{2}+p_{z}^{2})+r^{2}m^{2}u}%
{m^{3}\left(  r^{2}-x^{2}-y^{2}\right)  }\approx0. \label{for}%
\end{align}
Eqs. (\ref{thi}) and (\ref{for}) determine the Lagrangian multipliers
$\lambda$ and $u$ respectively. With introduction of the Dirac bracket instead
of the Poisson one for the canonical variables $A$ and $B,$\cite{dirac2}%
\begin{equation}
\left\{  A,B\right\}  _{D}=\left\{  A,B\right\}  -\left\{  A,\varphi_{\alpha
}\right\}  C_{\alpha\beta}^{-1}\left\{  \varphi_{\beta},B\right\}  ,
\end{equation}
which the matrix elements $C_{\alpha\beta}\left(  \alpha,\beta=1,2,3,4\right)
$ is defined by,%
\begin{equation}
C_{\alpha\beta}=\left\{  \varphi_{\alpha},\varphi_{\beta}\right\}  ,
\end{equation}
the positions $x_{i}$ and the momenta $p_{i}$ satisfy following Dirac bracket,
\cite{1968,1983,1985,1990,1992,1996-1,1996-2,1998,2000,2009}
\begin{align}
\left\{  x_{i},x_{j}\right\}  _{D}  &  =0,\left\{  x_{i},p_{j}\right\}
_{D}=\delta_{ij}-\frac{x_{i}x_{j}}{r^{2}},\nonumber\\
\left\{  p_{i},p_{j}\right\}  _{D}  &  =-\frac{1}{r^{2}}(x_{i}p_{j}-x_{j}%
p_{i}), \label{xp}%
\end{align}
and other Dirac brackets between $x_{i}$ and $p_{j}$ vanish. The equation of
motion is in general%
\begin{equation}
\dot{f}=\left\{  f,H_{p}\right\}  _{D}, \label{geneqm}%
\end{equation}
from which we have for $x_{i}$ and $p_{i}$ respectively,%
\begin{align}
\dot{x}_{i}  &  =\left\{  x_{i},H_{p}\right\}  _{D}=\frac{p_{i}}{m}%
,\label{xh}\\
\dot{p}_{i}  &  =\left\{  p_{i},H_{p}\right\}  _{D}=-\frac{x_{i}p_{i}^{2}%
}{mr^{2}}. \label{ph}%
\end{align}
Note that in these calculations (\ref{geneqm}), (\ref{xh}) and (\ref{ph})
where the constraints are of second-class, we need to deal with $H$ instead of
$H_{p}$ for we have,%
\begin{equation}
\dot{f}=\left\{  f,H_{p}\right\}  _{D}=\left\{  f,H\right\}  _{D}.
\end{equation}

In quantum mechanics, the Hamiltonian is, \cite{jk,dacosta}
\begin{align}
H  &  =-\frac{\hbar^{2}}{2m}\nabla^{2}=-\frac{\hbar^{2}}{2m}\frac{1}{\sqrt{g}%
}\partial_{\mu}\sqrt{g}g^{\mu\nu}\partial_{\upsilon}+V_{gp}\nonumber\\
&  =-\frac{\hbar^{2}}{2m}(\frac{r^{2}-x^{2}}{r^{2}}\frac{\partial^{2}%
}{\partial x^{2}}-\frac{2xy}{r^{2}}\frac{\partial}{\partial x}\frac{\partial
}{\partial y}-\frac{2x}{r^{2}}\frac{\partial}{\partial x}-\frac{2y}{r^{2}%
}\frac{\partial}{\partial y}+\frac{r^{2}-y^{2}}{r^{2}}\frac{\partial^{2}%
}{\partial y^{2}})+V_{gp}\text{ } \label{h}%
\end{align}
where the factor $g\equiv\det(g_{\mu\upsilon})$ is the determinant of the
matrix $g_{\mu\upsilon}$, and the geometric potential $V_{gp}=0$ for
$M^{2}-K=0$ on sphere. The quantum commutator $\left[  A,B\right]  $ of two
variables $A$ and $B$ is attainable by direct correspondence of the Dirac
bracket as $\left[  A,B\right]  /i\hbar\rightarrow\left\{  A,B\right\}  _{D}$,
and the fundamental commutators are:%
\begin{align}
\left[  x_{i},x_{j}\right]   &  =0,\label{dxx}\\
\left[  x_{i},p_{j}\right]   &  =i\hbar\left(  \delta_{ij}-\frac{x_{i}x_{j}%
}{r^{2}}\right)  ,\label{dxp}\\
\left[  p_{i},p_{j}\right]   &  =-\frac{i\hbar}{r^{2}}(x_{i}p_{j}-x_{j}p_{i}),
\label{dpp}%
\end{align}
There is no operator ordering problem in the right-hand side of Eq.
(\ref{dpp}) because the commutator must satisfy the Jacobian identity. It is
easily to verify that operators $p_{i}$ (\ref{px})-(\ref{pz}) satisfy
relations (\ref{dxx})-(\ref{dpp}). The second category of the fundamental
commutators is given by quantization of (\ref{xh}) and (\ref{ph})%
\begin{align}
\left[  x_{i},H\right]   &  =i\hbar\frac{p_{i}}{m},\label{qxh}\\
\left[  p_{i},H\right]   &  =-i\hbar\frac{x_{i}H+Hx_{i}}{mr^{2}}. \label{pH}%
\end{align}
Strikingly, the geometric momentum (\ref{px})-(\ref{pz}) satisfies all
commutators (\ref{dxp})-(\ref{pH}) above, not only (\ref{dxp})-(\ref{dpp}). As
pointed out in Ref. \cite{liuPRA}, the usual canonical momentum $p_{\theta}$
violates the fundamental commutator (\ref{pH}).

\section{The self-adjointness of the Geometric Momentum}

By a self-adjoint operator, we mean that all eigenvalues of it are real, and
eigenfunctions corresponding to distinct eigenvalues are mutually orthogonal
and they form a complete set. But direct demonstration of the self-adjointness
of the geometric momentum (\ref{px})-(\ref{pz}) is relatively difficult. With
variable transform $(x,y,z)\rightarrow(r,\theta,\varphi)$ with $\theta
\in(0,\pi),\varphi\in(0,2\pi)$ as,
\begin{equation}
\mathbf{r}\left(  \theta,\varphi\right)  =\left(  r\sin\theta\cos\varphi
,r\sin\theta\sin\varphi,r\cos\theta\right)
\end{equation}
being made, the task becomes easy. The geometric momentum operators in terms
of ($\theta,\varphi$) take following forms, \cite{liu07,liu03,liu10}%
\begin{align}
p_{x}  &  =-\frac{i\hbar}{r}\left(  \cos\theta\cos\varphi\frac{\partial
}{\partial\theta}-\frac{\sin\varphi}{\sin\theta}\frac{\partial}{\partial
\varphi}-\sin\theta\cos\varphi\right)  ,\\
p_{y}  &  =-\frac{i\hbar}{r}\left(  \cos\theta\sin\varphi\frac{\partial
}{\partial\theta}-\frac{\cos\varphi}{\sin\theta}\frac{\partial}{\partial
\varphi}-\sin\theta\sin\varphi\right)  ,\\
p_{z}  &  =-\frac{i\hbar}{r}\left(  -\sin\theta\frac{\partial}{\partial\theta
}-\cos\theta\right)  .
\end{align}
Their complete solutions to eigenvalue equations $p_{j}(\theta,\varphi
)\psi_{p_{j}}(\theta,\varphi)=p_{j}\psi_{p_{j}}(\theta,\varphi)$ can be easily
determined. To note that the eigenvalues $p_{j}$ on the right hand side of
these equations differ from the operators $p_{j}(\theta,\varphi)$ on the left
hand side. Explicitly, the solutions are, \cite{liu12}
\begin{align}
\psi_{p_{x}}\left(  \theta,\varphi\right)   &  =f_{x}\left(  \frac
{1+\sin\theta\cos\varphi}{1-\sin\theta\cos\varphi}\right)  ^{irp_{x}/2\hbar
}\frac{\sqrt{\cos\theta\sin\theta\sin\varphi}}{1-\sin^{2}\theta\cos^{2}%
\varphi},\label{f1}\\
\psi_{p_{y}}\left(  \theta,\varphi\right)   &  =f_{y}\left(  \frac
{1+\sin\theta\sin\varphi}{1-\sin\theta\sin\varphi}\right)  ^{irp_{y}/2\hbar
}\frac{\sqrt{\cos\theta\sin\theta\cos\varphi}}{1-\sin^{2}\theta\sin^{2}%
\varphi},\label{f2}\\
\psi_{p_{x}}\left(  \theta,\varphi\right)   &  =C_{z}\left(  \cot\frac{\theta
}{2}\right)  ^{irp_{z}/\hbar}\frac{1}{\sin\theta}, \label{f3}%
\end{align}
where $f_{x}$ and $f_{y\text{ }}$ are two arbitrary functions of the same
variable $\tan\theta\sin\varphi$. In terms of variables $(x,y,z)$, we have
from (\ref{f1})-(\ref{f3}),%
\begin{align}
\psi_{p_{x}}\left(  x,y\right)   &  =f_{x}\left(  \frac{r+x}{r-x}\right)
^{irp_{x}/2\hbar}\frac{r}{r^{2}-x^{2}}\sqrt{y\sqrt{r^{2}-x^{2}-y^{2}}},\\
\psi_{p_{y}}\left(  x,y\right)   &  =f_{y}\left(  \frac{r+y}{r-y}\right)
^{irp_{y}/2\hbar}\frac{r}{r^{2}-y^{2}}\sqrt{x\sqrt{r^{2}-x^{2}-y^{2}}},\\
\psi_{p_{z}}\left(  x,y\right)   &  =C_{z}\left(  \frac{r+\sqrt{r^{2}%
-x^{2}-y^{2}}}{\sqrt{x^{2}+y^{2}}}\right)  ^{irp_{z}/\hbar}\frac{r}%
{\sqrt{x^{2}+y^{2}}},
\end{align}
where $f_{x}$ and $f_{y\text{ }}$ are two arbitrary functions of variable
$y/\sqrt{r^{2}-x^{2}-y^{2}}$. One can then verify that $p_{j}(x,y)\psi_{p_{j}%
}(x,y)=p_{j}\psi_{p_{j}}(x,y)$ are satisfied with geometric momentum of form
(\ref{px})-(\ref{pz}).

\section{Remarks and summary}

Two dimensional surface can be considered as three dimensional shell whose
thickness is negligible in comparison with the dimension of the whole system.
We can study the quantum mechanics on surface by first formulating it in the
bulk, and then taking the limit of vanishing thickness, the gradient operator
and the Laplace operator originally defined in flat space converges to the
geometric ones. The presence of the geometric momentum and geometric potential
well reflects the Dirac's penetrating insight into the canonical quantization.

On the two dimensional sphere embedded\ in three dimensional flat space, the
geometric momentum in the Monge paramerization is extensively explored in this
paper. The apparent commutator $[x_{i},p_{j}]=i\hbar\delta_{ij}$ does not hold
true any more, and we must resort to the Dirac's theory on second-class
constrained motion. The correct results turn out to be $\left[  x_{i}%
,p_{j}\right]  =i\hbar\left(  \delta_{ij}-x_{i}x_{j}/r^{2}\right)  $. This
geometric momentum is geometric invariant under parameters transformation, and self-adjoint.

\begin{acknowledgments}
This work is financially supported by National Natural Science Foundation of
China under Grant No. 11175063.
\end{acknowledgments}

\end{document}